\documentclass[12pt]{article}

\ifx\pdfoutput\undefined
\usepackage[dvips,bookmarks]{hyperref}
\else
\usepackage{hyperref}
\fi
\hypersetup{colorlinks=false,bookmarksopen,bookmarksnumbered,citecolor=blue,
   pdfstartview=FitH}

\usepackage{latexsym}
\usepackage{amssymb,amsfonts,amsmath}
\usepackage{graphicx} 
\usepackage{indentfirst}
\usepackage{bbm}
\usepackage{amssymb}
\usepackage{verbatim}
\usepackage{amsmath, amsthm,amssymb}
\usepackage{mathrsfs}
\usepackage{hyperref}
\usepackage{amsfonts}
\usepackage{dsfont}

\parskip=\medskipamount

\newcommand {\cD}{{\cal D}}
\newcommand {\cE}{{\cal E}}
\newcommand {\cF}{{\cal F}}
\newcommand {\cG}{{\cal G}}
\newcommand {\cH}{{\cal H}}

\newcommand {\cJ}{{\cal J}}

\newcommand {\cN}{{\cal N}}

\newcommand {\cT}{{\cal T}}

\newcommand {\cV}{{\cal V}}
\newcommand {\cW}{{\cal W}}

\newcommand {\cY}{{\cal Y}}
\newcommand {\cZ}{{\cal Z}}

\def\a{\alpha}
\def\b{\beta}

\def\d{\delta}

\def\g{\gamma}
\def\G{\Gamma}

\def\p{\pi}
\def\q{\theta}

\def\s{\sigma}

\def\D{\Delta}
\def\F{\Phi}
\def\J{\Psi}
\def\L{\Lambda}
\def\O{\Omega}

\def\S{\Sigma}
\def\U{\Upsilon}
\def\X{\Xi}

\def\ri{{\rm i}}
\def\re{{\rm e}}

\newcommand{\ad}{{\dot{\alpha}}}                           
\newcommand{\bd}{{\dot{\beta}}}                            
\newcommand{\ve}{\varepsilon}                            
\newcommand{\cDB}{{\bar\cD}}                            

\newcommand{\hf}{\frac12}

\newcommand{\be}{\begin{equation}}
\newcommand{\ee}{\end{equation}}
\newcommand{\bea}{\begin{eqnarray}}
\newcommand{\eea}{\end{eqnarray}}
\newcommand{\non}{\nonumber}
\newcommand{\ba}{\begin{array}}
\newcommand{\ea}{\end{array}}

\newcommand{\1}{{\underline{1}}}
\newcommand{\2}{{\underline{2}}}

\newcommand{\dsC}{{\mathbb C}}

\newcommand{\bm}[1]{\mbox{\boldmath$#1$}}

\def\double #1{#1{\hbox{\kern-2pt $#1$}}}

\newcommand{\gd}{{\dot\g}}
\newcommand{\dd}{{\dot\d}}

\newcommand{\sba}{{\bar{\s}}}

\newcommand{\bsubeq}{\begin{subequations}}
\newcommand{\esubeq}{\end{subequations}}

\newcommand{\dalpha}{{\dot{\alpha}}}
\newcommand{\dbeta}{{\dot{\beta}}}

\newcommand{\N}{{\mathcal N}}

\newcommand{\rd}{\mathrm d}
\newcommand{\BCD}{{\bar\cD}}

\numberwithin{equation}{section}

\begin{document}

\begin{titlepage}
\begin{flushright}
July, 2013
\\
\end{flushright}
\vspace{5mm}

\begin{center}
{\Large \bf 
Super-Weyl anomalies in \mbox{$\bm{ \cN=2}$} supergravity  \\
and  (non)local
effective actions }
\\ 
\end{center}

\begin{center}

{\bf
Sergei M. Kuzenko} \\
\vspace{5mm}

\footnotesize{
{\it School of Physics M013, The University of Western Australia\\
35 Stirling Highway, Crawley W.A. 6009, Australia}}  
~\\
\vspace{2mm}

\end{center}

\begin{abstract}
\baselineskip=14pt
Using the formulation for $\cN=2$ conformal supergravity in SU(2) superspace, 
we define super-Weyl (or superconformal) anomalies and construct two types of 
nonlocal effective actions that generate these anomalies.
We also present the local Wess-Zumino 
action for spontaneously broken $\cN=2$ superconformal symmetry,  
with the Goldstone supermultiplet identified with a reduced chiral superfield
 containing the dilaton and the axion among its components.  
\end{abstract}

\vfill

\vfill
\end{titlepage}

\newpage
\renewcommand{\thefootnote}{\arabic{footnote}}
\setcounter{footnote}{0}




\section{Introduction}

Within the superconformal approach to $\cN=2$ 
locally supersymmetric theories in four dimensions
(see, e.g., \cite{FVP} for a pedagogical review), 
Poincar\'e or anti-de Sitter supergravity
is realized as conformal supergravity coupled to two compensators, 
one of which is invariably a vector multiplet.
If one makes use of  the superspace formulation for $\cN=2$ 
conformal supergravity developed in \cite{KLRT-M1}, 
 any supergravity-matter action is required to be invariant under 
arbitrary super-Weyl transformations generated by a covariantly 
chiral scalar parameter $\s$.\footnote{Although there exist 
two alternative superspace formulations for $\cN=2$ conformal supergravity
\cite{Howe,Butter11}, 
the approach described in \cite{KLRT-M1} is intimately related to
the structure of superconformal transformations in $\cN=2$ Minkowski superspace
as described in \cite{KT} (see also \cite{Park}).}
Given a superconformal field theory coupled to background $\cN=2$ supergravity, 
its classical action must be independent of the compensators. At the quantum level, 
however, integrating out the matter fields in such a theory 
results in the breakdown of the Weyl and local U$(1)_R$ symmetries.
There are two equivalent ways to describe this anomaly in terms of 
a nonlocal effective action that corresponds 
to the theory under consideration. Firstly, the effective action  $\G$ 
can be chosen in such a way that it does not depend explicitly on the compensating vector multiplet, but is not super-Weyl invariant, $\d_\s \G \neq 0$.  
The second option is that  the effective action $\widetilde{\G}$ is super-Weyl invariant, 
$\d_\s \widetilde{\G} =0$, but it depends explicitly on 
the compensating vector multiplet.
It turns out that $\G$ is obtained from $\widetilde{\G}$ by choosing an appropriate 
super-Weyl gauge. 

Recently, new $\cN=2$ locally supersymmetric 
higher-derivative invariants have been constructed that include 
the $\cN=2$ supersymmetric extension of the Gauss-Bonnet topological invariant
\cite{BdeWKL}. This work has actually provided the missing  tool 
one needs in order to give a simple and efficient superspace definition 
of the $\cN=2$ super-Weyl anomaly as well as to construct 
a nonlocal effective action that generates the anomaly. 
Moreover, it makes it possible to 
derive an $\cN=2$ supersymmetric extension of the local dilaton 
effective action introduced in \cite{SchT}.

This paper is organized as follows.
In section 2 we define consistent $\cN=2$ super-Weyl anomalies
and construct two types of nonlocal effective actions that generate these anomalies.
The local dilaton effective action is introduced in section 3.
Several implications of the results obtained are discussed in section 4.
Appendix A contains a  brief summary of the formulation for $\cN=2$ conformal 
supergravity  \cite{KLRT-M1} in SU(2) superspace.
Appendix B contains the expressions for the vector and tensor supergravity 
compensators in terms of prepotentials. 

\section{Nonlocal effective action}

Consider a superconformal field theory coupled to 
$\cN=2$ Poincar\'e or anti-de Sitter supergravity. The classical action 
of such a theory is invariant under the super-Weyl transformations, and it 
is independent of the supergravity compensators. 
In other words, the superconformal field theory couples to the Weyl multiplet. 
For concreteness, we use the supergravity formulation 
in which the second compensator is an improved tensor multiplet \cite{deWPV}.

In the quantum theory, integrating out the matter fields leads to an effective action 
that is no longer a functional of the Weyl multiplet only. 
As mentioned in section 1, 
there are two equivalent ways to describe such a functional 
with broken super-Weyl invariance. 
Let us first consider 
the realization in which the effective action $\G$ does not depend explicitly 
on the vector compensator, but is not invariant under the super-Weyl transformations.

In general, the super-Weyl variation of the effective action $\G$ has the form
\bea
\d_\s \G = (c-a) \int \rd^4x\, \rd^4\q\, \cE\, \s W^{\a\b}W_{\a\b} 
+ a \int \rd^4x\, \rd^4\q\, \cE\, \s \X ~+~{\rm c.c.} ~,
\label{2.1}
\eea
for some  anomaly coefficients $a$ and $c$. 
Here $\cE$ denotes the chiral density
(see, e.g.,  \cite{KT-M09} for more details), $W_{\a\b} =W_{\b\a} $  is the covariantly chiral
super-Weyl tensor, and $\X$  denotes the following composite scalar \cite{BdeWKL}:
\bea
\X :=  \frac{1}{6}  \bar{\cD}^{ij} \bar S_{ij}+  \bar S^{ij} \bar S_{ij}
+ \bar Y_{\dalpha \dbeta} \bar Y^{\dalpha \dbeta}~, \qquad
\cDB_{ij}:=\cDB_{\ad(i}\cDB_{j)}^\ad~.
\eea   
The torsion superfields $S_{ij}$, $W_{\a\b}$ and $Y_{\a\b}$ and their conjugates
$\bar S^{ij} $, $\bar W_{\ad \bd}$  and $\bar Y_{\ad \bd}$ 
are defined in  Appendix \ref{grimmspace}. 
The fundamental properties of $\X$ are as follows \cite{BdeWKL}: \\
(i) $\X$ is covariantly chiral, 
\begin{subequations}
\bea
\bar \cD^\ad_i \X=0~;
\eea
 (ii) the super-Weyl transformation of $\X$  is (see \eqref{chiral-pr} for the definition of 
 $\bar \D$)
 \bea
 \d_\s \X = 2\s \X -2 \bar \D \bar  \s ~;
\label{sW-X}
 \eea
(iii) the functional 
\bea
-\int \rd^4x\, \rd^4\q\, \cE\, \Big\{ W^{\a\b}W_{\a\b}
-\X\Big\} 
\label{topological}
\eea
\end{subequations}
is a topological invariant, which is related to the difference 
of the Gauss-Bonnet and Pontryagin invariants. 
The chirality of $\X$ is quite a nontrivial property which follows from
eqs. \eqref{acr2} and \eqref{A.5}. The super-Weyl transformation law \eqref{sW-X}
may be derived with the aid of  the equations \eqref{A.7b}, \eqref{super-Weyl-S} 
and \eqref{super-Weyl-Y}.

In the transformation law \eqref{sW-X}, 
  $\bar{\D}$ denotes the chiral projection operator \cite{KT-M09,Muller}
\bea
\bar{\D}
&=&\frac{1}{96} \Big((\cDB^{ij}+16\bar{S}^{ij})\cDB_{ij}
-(\cDB^{\ad\bd}-16\bar{Y}^{\ad\bd})\cDB_{\ad\bd} \Big)
\non\\
&=&\frac{1}{96} \Big(\cDB_{ij}(\cDB^{ij}+16\bar{S}^{ij})
-\cDB_{\ad\bd}(\cDB^{\ad\bd}-16\bar{Y}^{\ad\bd}) \Big)~,
\label{chiral-pr}
\eea
with $\cDB^{\ad\bd}:=\cDB^{(\ad}_k\cDB^{\bd)k}$.
Its main properties are the following:
for any super-Weyl inert scalar $U$
\begin{subequations} 
\bea
{\bar \cD}^{\ad}_i \bar{\D} U &=&0~, \\
\d_\s U = 0 \quad \Longrightarrow \quad 
\d_\s \bar \D U &=& 2\s \bar \D U~,  \label{2.5b}\\
\int \rd^4 x \,{\rm d}^4\q\,{\rm d}^4{\bar \q}\,E\, U
&=& \int {\rm d}^4x \,{\rm d}^4 \q \, \cE \, \bar{\D} U ~.
\label{chiralproj1} 
\eea
\end{subequations}
The super-Weyl invariance of the second term in \eqref{topological} follows 
from the relations \eqref{sW-X} and  \eqref{chiralproj1} 
in conjunction with the identity 
\bea
{\bar \cD}^{\ad}_i  \s=0
\quad \Longrightarrow \quad 
\int \rd^4 x \,{\rm d}^4\q\,{\rm d}^4{\bar \q}\,E\, \s=0~,
\label{2.6}
\eea
for any covariantly chiral scalar $\s$. 
    
One can check that the super-Weyl variation \eqref{2.1}
obeys the Wess-Zumino consistency condition \cite{WZ71}
\bea
(\d_{\s_1} \d_{\s_2} -\d_{\s_2} \d_{\s_1} ) \G=0~.
\eea
This property guarantees the existence of $\G$. 
    
To construct $\G$ explicitly,  
we introduce two scalar Green's functions $G_{+-} (z, z') $ and $G_{-+} (z, z') $ 
that are related to each other by the rule 
\bea
G_{+-} (z, z') = G_{-+} (z', z)
\eea
and obey the following conditions: \\
(i)   the two-point function $G_{-+} (z, z') $ is covariantly antichiral
in $z$ and chiral in $z'$, 
\bea
\cD^i_\a  G_{-+} (z, z') =0~, \qquad \bar \cD'{}_i^\ad  G_{-+} (z, z') =0~;
\eea
(ii) the two-point function $G_{-+} (z, z') $ satisfies the differential equation
\bea
\bar \D G_{-+} (z, z') = \d_+ (z,z')~ .
\label{Green}
\eea
Here we have used the  chiral delta-function 
\bea
 \d_+ (z,z'):= \bar \D\Big\{  E^{-1} \d^4(x-x') 
 \d^4 (\q-\q') \d^4 (\bar \q -\bar \q') \Big\} = \d_+(z',z)~,
 \eea
which is covariantly chiral with respect to each of its arguments,
\bea
\bar \cD^\ad_i  \d_+ (z,z')=0~, \qquad \bar \cD'{}^\ad_i  \d_+ (z,z')=0~.
\eea
Its main property is 
\bea
\J(z) =  \int {\rm d}^4x' \,{\rm d}^4 \q' \, \cE' \,\d_+(z,z')\, \J(z')~,
\qquad \bar \cD^\ad_i \J=0~,
\eea
for any covariantly chiral scalar $\J$.
Under a finite super-Weyl transformation,  
the full superspace delta-function $\d(z,z')= E^{-1} \d^4(x-x') 
 \d^4 (\q-\q') \d^4 (\bar \q -\bar \q')$ is inert, 
which implies that the chiral delta-function 
$ \d_+ (z,z')$  changes as follows:
\bea
 \d_+ (z,z') \to \re^{2\s}  \d_+ (z,z')~.
\label{sW-delta-function}
 \eea
${}$From the relations \eqref{2.5b}, \eqref{Green}
 and \eqref{sW-delta-function} it follows that the Green's functions $G_{-+} (z, z') $
and  $G_{+-} (z, z') $ are  super-Weyl inert. 
    
 Using the above relations allows us to determine a
 nonlocal effective action generating the super-Weyl anomaly \eqref{2.1}. It is 
 \bea
 \G&=&  -\hf (c-a)   \int {\rm d}^4x \,{\rm d}^4 \q \, \cE 
  \int {\rm d}^4x' \,{\rm d}^4 \q' \, \cE' \, W^{\a\b}(z)W_{\a\b} (z) G_{+-} (z,z') \bar \X (z')
  ~+~{\rm c.c.} \non \\
  && -\hf a \int {\rm d}^4x \,{\rm d}^4 \q \, \cE 
  \int {\rm d}^4x' \,{\rm d}^4 \q' \, \cE' \, \X (z) G_{+-} (z,z') \bar \X (z')~.
 \label{2.14}
  \eea
 This nonlocal effective action may be compared with the analogous results   
 in $\cN=1$ supergravity \cite{BK88,BK2013}.  
    
 In the superconformal approach to $\cN=2$ locally supersymmetric theories,
any supergravity-matter system is described by a super-Weyl invariant 
action functional that in general depends on  two compensators, one of which 
is always a vector multiplet.  To construct a super-Weyl invariant extension,
$\widetilde{\G} $, of the effective action \eqref{2.14}, we only need 
to take into account  the compensating vector multiplet  described 
by a covariantly chiral field strength $\cW$ and its conjugate $\bar \cW$, 
 \bea
{\bar \cD}^{\ad }_i \cW =0~,
\label{2.15}
\eea
subject to  the constraint \cite{KLRT-M1}
\bea
\Big(\cD^{ij}+4S^{ij}\Big)\cW
&=&
\Big(\cDB^{ij}+ 4\bar{S}^{ij}\Big)\bar{\cW}~,
\label{vectromul}
\eea
which reduces to that given in \cite{GSW} in the rigid supersymmetric limit.
In order to comply with the compensator interpretation, 
the field strength has to be nowhere vanishing, $\cW \neq 0$.
The constraints \eqref{2.15} and \eqref{vectromul} define a 
{\it reduced chiral} superfield.
Under the super-Weyl transformation,   $\cW$ varies as \cite{KLRT-M1}
\bea
\d_{\s} \cW = \s \cW~.
\label{Wsuper-Weyl}
\eea
The super-Weyl invariant extension of $\G$ is
\bea
\widetilde{\G} =\G  &-&\left[ \int \rd^4x\, \rd^4\q\, \cE\,  \Big\{ (c-a)\, W^{\a\b}W_{\a\b} 
+ a\,  \X \Big\}\ln \cW   ~+~{\rm c.c.} \right]
\non \\
&  - &2 a \int \rd^4 x \,{\rm d}^4\q\,{\rm d}^4{\bar \q}\,E\, \ln \cW \ln \bar \cW~.
\label{2.17}
 \eea
 It is easy to see that 
 \bea
 \d_\s \widetilde{\G} =0~.
 \eea   
 One may always choose a super-Weyl gauge $\cW=1$ in which the effective action 
   $ \widetilde{\G} $ reduces to   the expression
   \eqref{2.14}.\footnote{In case that $\cW = \rm const$,
 the second line of \eqref{2.17} vanishes, due to \eqref{2.6},
 for the choice $\s ={\rm const}$.} 
Since the vector multiplet described by $\cW$ and $\bar \cW$ is
 just a compensator, the super-Weyl symmetry  is not anomalous in this approach,
 in accordance with the general analysis given in \cite{deWG}.

\section{Dilaton effective action}

We now present a local Wess-Zumino action, $S_{\rm D}$, 
that reproduces the $\cN=2$ super-Weyl anomaly \eqref{2.1}.  
 It may be obtained by following the Wess-Zumino construction 
of the local action via integration of the anomaly \cite{WZ71}. 
In complete analogy with the $\cN=1$ 
case studied by Schwimmer and Theisen \cite{SchT}, 
our approach makes use of an off-shell dilaton supermultiplet
that contains the dilaton and the axion among its component fields.
We choose it to be the $\cN=2$  vector multiplet
described by a covariantly chiral field strength $\cZ$ and its conjugate $\bar \cZ$.     
The field strength $\cZ$ is a reduced chiral superfield obeying  the constraints   
 \eqref{2.15} and \eqref{vectromul}, with $\cW \to \cZ$. In addition,   
$\cZ$ is required to be nowhere vanishing, $\cZ \neq 0$. 
The  super-Weyl transformation of $\cZ$ written in the form 
\bea
\ln \cZ ~\to ~ \ln \cZ ' = \ln \cZ + \s
\label{3.1}
\eea
 tells us that $\ln \cZ $  is a Goldstone multiplet of spontaneously 
 broken super-Weyl symmetry. 
    
The local effective action is
\bea
S_{\rm D} [\cZ, \bar \cZ ] &=& \frac{1}{4} f^2\int \rd^4x\, \rd^4\q\, \cE\, \cZ^2 
+
\int \rd^4x\, \rd^4\q\, \cE\,  \Big\{  (c-a)\, W^{\a\b}W_{\a\b} 
+ a\,  \X \Big\}\ln \cZ   ~+~{\rm c.c.} \non \\
&&  +2 a \int \rd^4 x \,{\rm d}^4\q\,{\rm d}^4{\bar \q}\,E\, \ln \cZ \ln \bar \cZ~.
\label{3.2}
\eea
Its super-Weyl variation is
\bea
\d_\s S_{\rm D}[ \cZ, \bar \cZ ]  = (c-a) \int \rd^4x\, \rd^4\q\, \cE\, \s W^{\a\b}W_{\a\b} 
+ a \int \rd^4x\, \rd^4\q\, \cE\, \s \X ~+~{\rm c.c.} 
\eea
This variation coincides with \eqref{2.1} and   is independent of the dilaton supermultiplet.
The kinetic term in the first line of \eqref{3.2} is super-Weyl invariant. This term is added by hand.

In the flat-superspace limit, the functional in the second line of \eqref{3.2} 
reduces to the unique superconformal $F^4$ term discussed in \cite{deWGR,DS,BKT}.

The super-Weyl invariant extension   of $S_{\rm D} $ is 
\bea
\widetilde{S}_{\rm D} [\cZ, \bar \cZ ] &=& \frac{1}{4} f^2\int \rd^4x\, \rd^4\q\, \cE\, \cZ^2 
+
\int \rd^4x\, \rd^4\q\, \cE\,  \Big\{  (c-a)\, W^{\a\b}W_{\a\b} 
+ a\,  \X \Big\}\ln \frac{\cZ}{\cW}    ~+~{\rm c.c.} \non \\
&&  +2 a \int \rd^4 x \,{\rm d}^4\q\,{\rm d}^4{\bar \q}\,E\, 
\Big\{ 
\ln \cZ \ln \bar \cZ  - \ln \cW \ln \bar \cW \Big\} ~.
\eea
This action reduces to \eqref{3.2} in the super-Weyl gauge $\cW=1$. 
 
In the quantum theory, integrating out the dilaton supermultiplet leads to
a nonlocal  effective action depending only on the supergravity fields. 
By construction, this action reproduces the super-Weyl anomaly. 
At tree level we can approximate $\cZ$ by a chiral superfield $\mathfrak Z$
 obeying the equation 
\bea
\big(\cD_{ij}+4S_{ij}\big){\mathfrak Z} =0~,
\eea
 which is the equation of motion corresponding to the kinetic term in \eqref{3.2}. 
 This equation has a simple geometric interpretation if we recall 
 the finite super-Weyl transformation of the torsion tensor $S_{ij}$ \cite{KT-M-ads},
 \bea
 S_{ij} ~\to ~ S'_{ij} = \frac{1}{4} \re^{\s + \bar \s} \big(\cD_{ij}+4S_{ij}\big) \re^{-\s}~.
 \eea
 Thus the chiral parameter $\S$ defined by 
 ${\mathfrak Z} = \re^{- \S } $ generates a super-Weyl transformation such that $ S'_{ij} =0$.
 This transformation is unique under natural boundary conditions, which implies that 
 $\mathfrak Z$ is a well-defined functional of the supergravity fields. 
 Using the interpretation of $\mathfrak Z$ described, one may see that its super-Weyl 
 transformation law is 
 \bea
 {\mathfrak Z}~ \to~ {\mathfrak Z}'  = \re^\s{\mathfrak Z}~.
 \eea
 
 For the transformation law \eqref{3.1} and for the effective action \eqref{3.2},
it is not essential that  $\cZ$ obeys the Bianchi identity \eqref{vectromul}.
In principle, the constraint  \eqref{vectromul}
may be replaced with a more general off-shell condition 
\bea
\big(\cD^{ij}+4S^{ij}\big)\cZ
- \big(\cDB^{ij}+ 4\bar{S}^{ij}\big)\bar{\cZ} = \ri \,\cH^{ij} ~, \qquad
\cD^{(i}_\a \cH^{jk)} =  {\bar \cD}^{(i}_\ad \cH^{jk)} = 0~,
\eea
with $\cH^{ij}$   a  real ${\rm SU}(2)$ triplet.

\section{Discussion} 

It is of interest to discuss the effective action \eqref{2.17} 
in the context of the $\cN=2$ supercurrent multiplet introduced by Sohnius \cite{Sohnius}
in the super-Poincar\'e case (see also \cite{KT}). 
Within the superconformal setting for  $\cN=2$ supergravity, the supercurrent multiplet 
was defined in \cite{BK11}, and here we recall the key relations.
To start with, we have to recall  the prepotential structure of 
$\cN=2$ supergravity realized as the Weyl multiplet coupled to vector and tensor compensators. 
The Weyl multiplet can be described in terms of a real unconstrained prepotential 
$\cH$ (introduced at the linearized level in \cite{RT,HST})
which naturally originates (see \cite{KT} for a detailed derivation) 
within the harmonic-superspace approach to $\cN=2$ supergravity \cite{GIOS}.
A prepotential for the compensating vector multiplet can be chosen to be 
an unconstrained real SU(2) triplet $\cV^{ij} = \cV^{ji}$, 
$\overline{\cV^{ij}}= \cV_{ij} = \ve_{ik}\ve_{jl} \cV^{kl}$.
 The explicit expression for the chiral field strength $\cW$ in terms of the prepotential 
 is given by eq. \eqref{eq_WMezin}. Prepotentials for the compensating tensor multiplet
 can be chosen to be a covariantly chiral scalar $\J$ and its conjugate $\bar \J$.
 The explicit expression for the field strength, $\cG^{ij}$, of the tensor multiplet
 in terms of the prepotentials is given by eq. \eqref{eq_Gprepotential}.

The supercurrent conservation equation \cite{BK11} is
\bea
\label{N2current}
\frac{1}{4} (\BCD_{ij} + 4 \bar S_{ij}) \cJ = \cW \cT_{ij} - \cG_{ij} \cY~,
\eea
where the supercurrent $\cJ$ and the trace multiplets $ \cT_{ij} $ and $\cY$ are defined as
\bea
\cJ = \frac{{\bm  \d} S}{{ \bm\d}\cH }~, \qquad \cT_{ij} = \frac{{\d} S}{{\d} \cV^{ij} }~, 
\qquad \cY = \frac{{\d} S}{{\d} \J }~,
\eea
with $ {{\bm  \d} S}/{{ \bm\d}\cH }$ a covariantized variational derivative.
The superfields  $\cJ$ and $\cT_{ij}$ are real,  while $\cY$ is covariantly  chiral.
In addition, $\cY$ and $\cT_{ij}$ must obey the constraints
\begin{subequations}
\begin{gather}\label{eq_CurrentConstraints}
\cD_\alpha{}^{(k} \cT^{ij)} = \bar\cD_\dalpha{}^{(k} \cT^{ij)} = 0 ~,\\
(\cD^{ij} + 4 S^{ij})\cY = (\bar \cD^{ij} + 4 \bar S^{ij})\bar\cY~.
\end{gather}
\end{subequations}
These constraints are due to the property that the prepotential $\cV^{ij}$ and $\cJ$
are defined modulo the gauge transformations \eqref{pre-gauge1}
and \eqref{eq_PsiGauge} respectively. In the case of a superconformal field theory, 
both trace multiplets vanish, 
\bea
\cT^{ij} =0~, \qquad \cY=0~.
\eea

For the effective action \eqref{2.17},  we easily read off the trace multiplets
\bea
\langle \cT_{ij}\rangle &=& -\frac{1}{4} (\cD_{ij}+4S_{ij} ) 
\Big\{ (c-a) \frac{W^{\a\b}W_{\a\b} }{\cW}  
+a \frac{ \X+2\bar \D \ln \bar \cW} {\cW} \Big\} \,+\,{\rm c.c.}~, \qquad
\langle \cY \rangle =0~.~~~
\label{AST}
\eea
The effective action is independent of the tensor compensator, which implies 
$\langle \cY \rangle =0$.

It is instructive to compare the anomalous supertrace \eqref{AST} 
with that describing the super-Weyl anomalies in $\cN=1$ supergravity \cite{BPT,BK86}
(see \cite{BK} for a review)\footnote{Direct calculations of 
 the anomalous supertrace in concrete models \cite{BK86} produce 
an additional contribution to $\langle T\rangle$ which is proportional 
to $(\bar \cD^2 - 4R ) \cD^2 R $. It can  be removed by adding 
a local finite counterterm $\int \rd^4 x \,\rd^2 \q \,\rd^2 \bar \q \,E \, R\bar R$
to the effective  action.}
\bea
\langle T\rangle= (c -a) W^{\a\b\g} W_{\a\b\g} 
+ \frac{a}{4} (\bar \cD^2 - 4R ) (G^aG_a +2R \bar R)~,
\label{anomaly}
\eea
with $a$ and $c$ numerical parameters (containing a factor of $1/\p^2$).
Here $W_{\a\b\g}$ is the covariantly chiral super-Weyl tensor, 
and $G_a$, $R$ and $\bar R$ are the other torsion superfields
of the Wess-Zumino superspace geometry \cite{WZ,GWZ}.
The $a$-terms in \eqref{anomaly} constitute 
 the chiral density of the $\N=1$ topological invariant
\cite{FZ78} 
\bea
\int \rd^4 x \,\rd^2 \q \,\cE \,W^{\a\b\g} W_{\a\b\g} 
+ \int \rd^4 x \,\rd^2 \q \,\rd^2 \bar \q \,E \, (G^aG_a +2R \bar R)~.
\eea
The anomalous supertrace \eqref{anomaly} corresponds to the 
 conservation equation \cite{FZ}
\bea
\bar \cD^\ad T_{\a\ad} +\frac{2}{3} \cD_\a T = 0~, \qquad \bar \cD_\ad T =0~,
\label{4.8}
\eea
with $T_a= \bar T_a$ the supercurrent and $T$ the trace multiplet.
Unlike the $\cN=2$ supercurrent equation \eqref{N2current},
the conservation law \eqref{4.8} does not involve any compensator. 
The point is that \eqref{4.8} originates within the ordinary Wess-Zumino formulation 
for $\cN=1$ supergravity.
One can develop a superconformal  extension of this formulation  
 that makes use of a covariantly chiral scalar compensator $\F$, $\bar \cD_\ad \F=0$, 
 and any supergravity-matter action is invariant under super-Weyl transformations.
 From the point of view of such a formulation, eq. \eqref{4.8} 
 corresponds to a super-Weyl gauge $\F=1$. Without imposing this gauge condition,
 the supercurrent equation \eqref{4.8} has  to be replaced with a more general equation given in \cite{Butter:2010hk}. Then, the anomalous supertrace \eqref{anomaly} gets modified to include an explicit dependence on $\F$.

The $\cN=1$ anomalous supertrace  \eqref{anomaly} was originally computed 
in \cite{BK86} for the scalar and vector multiplets coupled to $\cN=1$ supergravity.
It is an interesting open problem to compute the $\cN=2$
anomalous supertrace \eqref{AST} for  vector and hyper multiples in 
a manifestly supersymmetric setting.

The analysis given in section 2 implies the existence of 
the following super-Weyl inert covariantly chiral scalar: 
\bea
{\bm \J}^2 := \frac{ 1}  {\cW^2} \Big(  \bar \D \ln \bar \cW + \hf  \X \Big) ~, \qquad 
\bar \cD^\ad_i {\bm \J}^2 =0~, \qquad \d_\s {\bm \J}^2 =0~.
\label{4.5}
\eea
This is a curved superspace generalization of the conformal primary weight-zero 
chiral superfield introduced in \cite{BKT}. The superfield $\cW^2 {\bm \J}^2$ naturally 
originates within the approach developed in \cite{BdeWKL}. 
The chiral scalar ${\bm \J}^2 $ and its conjugate $\bar {\bm \J}^2 $
can be used to construct higher derivative super-Weyl invariants 
\begin{subequations}
\bea&& 
\int \rd^4 x \,{\rm d}^4\q\,\cE\,  \cW^2
\cF ({\bm \J}^2)~, \\
&& \int \rd^4 x \,{\rm d}^4\q\,{\rm d}^4{\bar \q}\,E\,  
\U ({\bm \J}^2, \bar {\bm \J}^2) ~, 
\eea
\end{subequations}
which generalize those introduced in \cite{BKT}. 
At the component level, such functionals generate $F^{2n} $ contributions,
with $n=3,4, \dots$, and $F$ the U(1) field strength. 
\\

\noindent
{\bf Acknowledgement:}\\
The author acknowledges helpful discussions with Daniel Butter and Stefan Theisen.
The generous hospitality of the Arnold Sommerfeld Center for Theoretical Physics 
at the University  for Munich, where this project was completed, 
is gratefully acknowledged. 
This work was supported in part by the Australian Research Council.

\appendix   

\section{Conformal supergravity} \label{grimmspace}
\allowdisplaybreaks

This appendix contains a summary of the formulation for $\cN=2$ conformal supergravity  \cite{KLRT-M1} in SU(2) superspace \cite{Grimm}.
A curved $\cN = 2$ superspace
parametrized by local coordinates 
$z^M = (x^m, \theta^\mu_\imath, \bar{\theta}_{\dot{\mu}}^\imath 
= (\theta_{\mu \imath})^* )$, where $m = 0, 1, ... \ , 3$, 
$\mu = 1, 2$. $\dot{\mu} = 1, 2$ and $\imath = \1, \2$.
The structure group is chosen to be $\rm SL(2, \dsC) \times SU(2)$, 
and the covariant derivatives $\cD_A = (\cD_a, \cD_\a^i, \bar \cD^\ad_i)$ read
\bea
\cD_A &=& E_A + \Phi_A{}^{kl} J_{kl}+ \hf \Omega_A{}^{bc} M_{bc} \non \\
		  &=& E_A + \Phi_A{}^{kl} J_{kl}+ \Omega_A{}^{\b\g} M_{\b\g} 
		  + \bar{\Omega}_A{}^{ \dot{\b} \dot{\g} } \bar M_{\dot{\b}\dot{\g}}~.
\eea
Here $M_{cd}$ and $J_{kl}$ are the generators of the Lorentz and SU(2) groups respectively, 
and $\O_A{}^{bc}$ and $\Phi_A{}^{kl}$ the corresponding connections. 
The action of the generators on the covariant derivatives are defined as:
\begin{subequations}
\bea
[M_{\a\b},  \cD_\g^i] &=& \ve_{\g (\a } \cD_{\b)}^i \ , \qquad [\bar M_{\ad \bd} , \bar \cD^i_\gd ] 
= \ve_{ \gd (\ad } \bar \cD^i_{\bd) i } ~, \\
\big[ J_{kl}, \cD_\a^i \big] &=& - \d^i_{(k} \cD_{\a l)} \ , \qquad 
[J_{kl}, \bar \cD^\ad_i] = - \ve_{i ( k} \bar \cD^\ad_{l)} \ .
\eea
\end{subequations}

The algebra of covariant derivatives is \cite{KLRT-M1}
\begin{subequations} \label{A.3}
\bea
\{\cD_\a^i,\cD_\b^j\}&=&\phantom{+}
4S^{ij}M_{\a\b}
+2\ve^{ij}\ve_{\a\b}Y^{\g\d}M_{\g\d}
+2\ve^{ij}\ve_{\a\b}\bar{W}^{\gd\dd}\bar{M}_{\gd\dd}
\non\\
&&
+2 \ve_{\a\b}\ve^{ij}S^{kl}J_{kl}
+4 Y_{\a\b}J^{ij}~,
\label{acr1} \\
\{\cDB^\ad_i,\cDB^\bd_j\}&=&
-4\bar{S}_{ij}\bar{M}^{\ad\bd}
-2\ve_{ij}\ve^{\ad\bd}\bar{Y}^{\gd\dd}\bar{M}_{\gd\dd}
-2\ve_{ij}\ve^{\ad\bd}{W}^{\g\d}M_{\g\d}
\non\\
&&
-2\ve_{ij}\ve^{\ad\bd}\bar{S}^{kl}J_{kl}
-4\bar{Y}^{\ad\bd}J_{ij}~,
\label{acr2} \\
\{\cD_\a^i,\cDB^\bd_j\}&=&
-2\ri\d^i_j(\s^c)_\a{}^\bd\cD_c
+4\d^{i}_{j}G^{\d\bd}M_{\a\d}
+4\d^{i}_{j}G_{\a\gd}\bar{M}^{\gd\bd}
+8 G_\a{}^\bd J^{i}{}_{j}~,
\label{acr3} 
\eea
\end{subequations}
see \cite{KLRT-M1} for the explicit expressions for  commutators
${[}\cD_a,\cD_\b^j{]}$ and ${[}\cD_a,\cDB^\bd_j{]}$.
Here the real four-vector $G_{\a \ad} $,
the complex symmetric  tensors $S^{ij}=S^{ji}$, $W_{\a\b}=W_{\b\a}$, 
$Y_{\a\b}=Y_{\b\a}$ and their complex conjugates 
$\bar{S}_{ij}:=\overline{S^{ij}}$, $\bar{W}_{\ad\bd}:=\overline{W_{\a\b}}$,
$\bar{Y}_{\ad\bd}:=\overline{Y_{\a\b}}$ 
are constrained by 
the Bianchi identities \cite{Grimm,KLRT-M1}.
   The latter  comprise 
 the  dimension-3/2 identities 
\begin{subequations}\label{A.5}
\bea
\cD_{\a}^{(i}S^{jk)}= {\bar \cD}_{\ad}^{(i}S^{jk)}&=&0~,
\label{S-analit}\\
\cDB^\ad_iW_{\b\g}&=&0~,\qquad\\
\cD_{(\a}^{i}Y_{\b\g)}&=&0~,\\
\cD_{\a}^{i}S_{ij}+\cD^{\b}_{j}Y_{\b\a}&=&0~, \\
\cD_\a^iG_{\b\bd}&=&
-{\frac14}\cDB_{\bd}^iY_{\a\b}
+\frac{1}{12}\ve_{\a\b}\cDB_{\bd j}S^{ij}
-{\frac14}\ve_{\a\b}\cDB^{\gd i}\bar{W}_{\bd\gd}~,
\eea
\end{subequations}
as well as the dimension-2 relation
\bea
\big( \cD^i_{(\a} \cD_{\b) i}
-4Y_{\a\b} \big) W^{\a\b}
&=& \big( \cDB_i^{( \ad }\cDB^{\bd ) i}
-4\bar{Y}^{\ad\bd} \big) \bar{W}_{\ad\bd}
~.
\label{dim-2-constr}
\eea

The algebra of covariant derivatives  \eqref{A.3} is invariant 
under the super-Weyl transformations  \cite{KLRT-M1}
\begin{subequations}
\bea
\d_{\s} \cD_\a^i&=&\hf\sba\cD_\a^i+(\cD^{\g i}\s)M_{\g\a}-(\cD_{\a k}\s)J^{ki}~, \\
\d_{\s} \cDB_{\ad i}&=&\hf\s\cDB_{\ad i}+(\cDB^{\gd}_{i}\sba)\bar{M}_{\gd\ad}
+(\cDB_{\ad}^{k}\sba)J_{ki}~, 
\label{A.7b}
 \\
\d_{\s} \cD_a&=&
\hf(\s+\sba)\cD_a
+{\frac{\ri}4}(\s_a)^\a{}_{\bd}(\cD_{\a}^{ k}\s)\cDB^{\bd}_{ k}
+{\frac{\ri}4}(\s_a)^{\a}{}_\bd(\cDB^{\bd}_{ k}\sba)\cD_{\a}^{ k} \non \\
&&\quad -{\frac12}\big(\cD^b(\s+\sba)\big)M_{ab}
~, 
\eea
\end{subequations}
with  the parameter $\s$ being an arbitrary covariantly chiral superfield, 
\bea
{\bar \cD}_{\ad i} \s=0~,
\eea
provided 
the  dimension-1 components of the torsion transform 
 as follows:
\begin{subequations}
\bea
\d_{\s} S^{ij}&=&\sba S^{ij}-{\frac14}\cD^{\g(i}\cD^{j)}_\g \s~, 
\label{super-Weyl-S} \\
\d_{\s} Y_{\a\b}&=&\sba Y_{\a\b}-{\frac14}\cD^{k}_{(\a}\cD_{\b)k}\s~,
\label{super-Weyl-Y} \\
\d_{\s} {W}_{\a \b}&=&\s {W}_{\a \b }~, \label{A.9c}\\
\d_{\s} G_{\a\bd} &=&
\hf(\s+\sba)G_{\a\bd} -{\frac{\ri}4}
\cD_{\a \bd} (\s-\sba)~.
 \label{super-Weyl-G} 
\eea
\end{subequations}
As is seen from \eqref{A.9c},
the covariantly chiral symmetric spinor $W_{\a\b}$ transforms homogeneously, 
and therefore it is a superfield extension of the Weyl tensor.   
    
\section{Compensators as gauge-invariant field strengths}\label{AppendixB}

In this appendix we recall expressions for the two supergravity compensators
in terms of prepotentials. 
    
The first compensator is identified with 
the reduced chiral superfield  $\cW $  and its conjugate $\bar \cW$. 
These superfields can be expressed in term of 
the curved-superspace extension \cite{n2_sugra_tensor} of Mezincescu's
prepotential \cite{Mezincescu},  $\cV^{ij}=\cV^{ji}$,
which is an unconstrained real SU(2) triplet. 
The expression for $\cW$ in terms of $\cV^{ij}$ \cite{n2_sugra_tensor} is
\begin{align}\label{eq_WMezin}
\cW = \frac{1}{4}\bar\Delta \Big({\cD}_{ij} + 4 S_{ij}\Big) \cV^{ij}~.
\end{align}
The prepotential is defined only up to gauge transformations \cite{n2_sugra_tensor}
\begin{align}
\delta \cV^{ij} &= \cD^{\alpha}{}_k \Lambda_\alpha{}^{kij}
     + \bar\cD_{\dalpha}{}_k \bar\Lambda^\dalpha{}^{kij}, \qquad
     \Lambda_\alpha{}^{kij} = \Lambda_\alpha{}^{(kij)}~,
     \qquad \bar\Lambda^\dalpha{}_{kij} := ( \Lambda_\alpha{}^{kij} )^*~,
\label{pre-gauge1}
\end{align}
with the gauge parameter $ \Lambda_\alpha{}^{kij} $ being completely arbitrary modulo 
the algebraic condition given.
The super-Weyl transformation of $V^{ij}$ is \cite{K-nsd} 
\bea
\d_\s V^{ij} = -(\s +\bar \s) V^{ij}~.
\eea 

 The second compensator is  
a tensor  (or {\it linear}) multiplet. It is  described
by a  real ${\rm SU}(2)$
triplet  $\cG^{ij}$, with the algebraic properies
$\cG^{ij}=\cG^{ji}$ and ${\bar \cG}_{ij}:=(\cG^{ij})^* = \cG_{ij}$,
subject to the constraints \cite{BS,SSW}
\bea
\cD^{(i}_\a \cG^{jk)} =  {\bar \cD}^{(i}_\ad \cG^{jk)} = 0~.
\label{2.5}
\eea
These constraints are solved \cite{HST,GS82,Siegel83,Muller86}
 in terms of a covariantly chiral
scalar prepotential $\Psi$ and its conjugate $\bar \J$ as follows: 
\begin{align}
\label{eq_Gprepotential}
\cG^{ij} = \frac{1}{4}\Big( \cD^{ij} +4{S}^{ij}\Big) \Psi
+\frac{1}{4}\Big( \cDB^{ij} +4\bar{S}^{ij}\Big){\bar \Psi}~, \qquad
{\bar \cD}_i^\ad \J=0~.
\end{align}
The prepotential is defined up to  gauge transformations of the form
\bea\label{eq_PsiGauge}
\d \J = {\rm i} \,\L ~, \qquad 
\Big(\cD^{ij}+4S^{ij}\Big) \L&=& 
\Big(\cDB^{ij} + 4\bar{S}^{ij}\Big)\bar{\L} ~,
\eea
with $\Lambda$ an arbitrary  reduced chiral superfield.
The super-Weyl transformation laws of $\cG^{ij}$ and $\J$ were given 
in \cite{KLRT-M1} and \cite{Kuz2008} respectively:
\bea
\d_\s \cG^{ij} = (\s +\bar \s) \cG^{ij}~, \qquad \d_\s \J = \s \J~.
\eea

\begin{footnotesize}

\end{footnotesize}


\begin{thebibliography}{66}
  
\bibitem{FVP} 
  D.~Z.~Freedman and A.~Van Proeyen,
 {\it Supergravity},
  Cambridge University Press,  Cambridge, 2012.

\bibitem{KLRT-M1}
S.~M.~Kuzenko, U.~Lindstr\"om, M.~Ro\v cek and G.~Tartaglino-Mazzucchelli,
``4D N=2 supergravity and projective superspace,'' 
JHEP {\bf 0809}, 051 (2008) [arXiv:0805.4683].

\bibitem{Howe}
P.~S.~Howe,
``A superspace approach to extended conformal supergravity,''
Phys.\ Lett.\  B {\bf 100} (1981) 389;
``Supergravity in superspace,'' Nucl.\ Phys.\  B {\bf 199} (1982) 309.

\bibitem{Butter11}
  D.~Butter,
  ``N=2 conformal superspace in four dimensions,''
  JHEP {\bf 1110} (2011) 030
  [arXiv:1103.5914 [hep-th]].
  
\bibitem{KT}
S.~M.~Kuzenko and S.~Theisen,
 ``Correlation functions of conserved currents in N = 2 superconformal
theory,''  Class.\ Quant.\ Grav.\  {\bf 17}, 665 (2000)  [hep-th/9907107]. 

\bibitem{Park}
  J.~H.~Park,
   ``Superconformal symmetry and correlation functions,''
  Nucl.\ Phys.\ B {\bf 559}, 455 (1999)   [hep-th/9903230].
 

  
  
 \bibitem{BdeWKL} 
 D. Butter, B. de Wit, S. M. Kuzenko and I. Lodato,
``New higher-derivative invariants in N=2  supergravity
and the Gauss-Bonnet term,'' 
 arXiv:1307.6546 [hep-th].


\bibitem{SchT} 
  A.~Schwimmer and S.~Theisen,
 ``Spontaneous breaking of conformal invariance and trace anomaly matching,''
  Nucl.\ Phys.\ B {\bf 847}, 590 (2011)
  [arXiv:1011.0696 [hep-th]].
  
 \bibitem{deWPV}
B.~de Wit, R.~Philippe and A.~Van Proeyen,
``The improved tensor multiplet in N = 2 supergravity,''
Nucl.\ Phys.\ B {\bf 219}, 143 (1983).

\bibitem{KT-M09}
  S.~M.~Kuzenko and G.~Tartaglino-Mazzucchelli,
 ``Different representations for the action principle in 4D N = 2 supergravity,''
  JHEP {\bf 0904} (2009) 007
  [arXiv:0812.3464 [hep-th]].
  
   \bibitem{Muller} M. M\"uller, {\it Consistent Classical Supergravity Theories},
(Lecture Notes in Physics, Vol. 336),
Springer, Berlin, 1989. 
 
\bibitem{WZ71} 
  J.~Wess and B.~Zumino,
  ``Consequences of anomalous Ward identities,''
  Phys.\ Lett.\ B {\bf 37}, 95 (1971).

\bibitem{BK88} 
I.~L.~Buchbinder and S.~M.~Kuzenko,
 ``Nonlocal action for supertrace anomalies in superspace of N=1 supergravity,''
Phys.\ Lett.\ B {\bf 202}, 233 (1988).  
  
\bibitem{BK2013} 
  D.~Butter and S.~M.~Kuzenko,
  ``Nonlocal action for the super-Weyl anomalies: A new representation,''
  arXiv:1307.1290 [hep-th].
 
\bibitem{GSW}
  R.~Grimm, M.~Sohnius and J.~Wess,
  ``Extended supersymmetry and gauge theories,''
  Nucl.\ Phys.\  B {\bf 133}, 275 (1978).
 
\bibitem{deWG} 
  B.~de Wit and M.~T.~Grisaru,
  ``Compensating fields and anomalies,''
 in  {\it Quantum Field Theory and Quantum Statistics, Vol. 2}, 
 I. A.  Batalin, C. J. Isham and G. A. Vikovisky (Eds.) Adam Hilger, Bristol, 
 1987, pp. 411--432.  
  

 



\bibitem{deWGR}
B.~de Wit, M.~T.~Grisaru and M.~Ro\v{c}ek,
``Nonholomorphic corrections to the one-loop 
N=2 super Yang-Mills action,''
Phys.\ Lett.\ B {\bf 374},  297 (1996) 
[arXiv:hep-th/9601115].

\bibitem{DS}
M.~Dine and N.~Seiberg,
``Comments on higher derivative operators in some 
SUSY field theories,''
Phys.\ Lett.\ B {\bf 409},  239 (1997) 
[arXiv:hep-th/9705057].

\bibitem{BKT}
I.~L.~Buchbinder, S.~M.~Kuzenko and A.~A.~Tseytlin,
``On low-energy effective actions in N = 2,4 superconformal 
theories  in  four dimensions,''
Phys.\ Rev.\ D {\bf 62},  045001 (2000) 
[arXiv:hep-th/9911221].



\bibitem{KT-M-ads}
  S.~M.~Kuzenko and G.~Tartaglino-Mazzucchelli,
  ``Field theory in 4D N=2 conformally flat superspace,''
  JHEP {\bf 0810}, 001 (2008)
  [arXiv:0807.3368 [hep-th]].


\bibitem{Sohnius}
  M.~F.~Sohnius,
  ``The multiplet of currents for N=2 extended supersymmetry,''
  Phys.\ Lett.\  B {\bf 81}, 8 (1979).
  


\bibitem{BK11} 
  D.~Butter and S.~M.~Kuzenko,
  ``N=2 AdS supergravity and supercurrents,''
  JHEP {\bf 1107}, 081 (2011)
  [arXiv:1104.2153 [hep-th]].


\bibitem{RT}
  V.~O.~Rivelles and J.~G.~Taylor,
 ``Linearised N=2 superfield supergravity,''
  J.\ Phys.\ A {\bf 15} (1982) 163.

\bibitem{HST}
  P.~S.~Howe, K.~S.~Stelle and P.~K.~Townsend,
  ``Supercurrents,''
  Nucl.\ Phys.\ B {\bf 192} (1981) 332.


\bibitem{GIOS}
  A.~S.~Galperin, E.~A.~Ivanov, V.~I.~Ogievetsky and E.~S.~Sokatchev,
  {\it Harmonic Superspace}, Cambridge University Press, 
  Cambridge, 2001.

\bibitem{BPT} 
  L.~Bonora, P.~Pasti and M.~Tonin,
``Cohomologies and anomalies in supersymmetric theories,''
  Nucl.\ Phys.\ B {\bf 252}, 458 (1985).  

\bibitem{BK86} 
  I.~L.~Buchbinder and S.~M.~Kuzenko,
  ``Matter superfields in external supergravity: Green functions, effective action and superconformal anomalies,''
  Nucl.\ Phys.\ B {\bf 274}, 653 (1986).
  
\bibitem{BK}
I.~L.~Buchbinder and S.~M.~Kuzenko,
{\it Ideas and methods of supersymmetry and supergravity: Or a walk through superspace},
Bristol, UK: IOP (1998) 656 p.



 \bibitem{WZ}
J.~Wess and B.~Zumino,
``Superfield Lagrangian for supergravity,''
Phys.\ Lett.\  B {\bf 74}, 51 (1978).

\bibitem{GWZ} 
  R.~Grimm, J.~Wess and B.~Zumino,
  ``A complete solution of the Bianchi identities in superspace,''
  Nucl.\ Phys.\ B {\bf 152}, 255 (1979).

\bibitem{FZ78}
  S.~Ferrara and B.~Zumino,
  ``Structure of linearized supergravity and conformal supergravity,''
  Nucl.\ Phys.\  B {\bf 134}, 301 (1978).
  
\bibitem{FZ}
 S.~Ferrara and B.~Zumino,
``Transformation properties of the supercurrent,''
Nucl.\ Phys.\  B {\bf 87}, 207 (1975).
  
\bibitem{Butter:2010hk} 
  D.~Butter,
  ``Conserved supercurrents and Fayet-Iliopoulos terms in supergravity,''
  arXiv:1003.0249 [hep-th].
  

\bibitem{Grimm}
 R.~Grimm,
``Solution of the Bianchi identities in SU(2) extended superspace with constraints,''
in {\it Unification of the Fundamental Particle Interactions}, 
S. Ferrara, J. Ellis and P. van Nieuwenhuizen (Eds.),
Plenum Press, New York, 1980, pp. 509-523.



\bibitem{n2_sugra_tensor}
  D.~Butter and S.~M.~Kuzenko,
  ``New higher-derivative couplings in 4D N = 2 supergravity,''
  JHEP {\bf 1103}, 047 (2011)
  [arXiv:1012.5153 [hep-th]].


\bibitem{Mezincescu}
  L.~Mezincescu,
  ``On the superfield formulation of O(2) supersymmetry,''
  Dubna preprint JINR-P2-12572 (June, 1979).

\bibitem{K-nsd} 
  S.~M.~Kuzenko,
  ``Nonlinear self-duality in N = 2 supergravity,''
  JHEP {\bf 1206}, 012 (2012)
  [arXiv:1202.0126 [hep-th]].


\bibitem{BS}
P.~Breitenlohner and M.~F.~Sohnius,
``Superfields, auxiliary fields, and tensor calculus for N=2 extended
supergravity,''
Nucl.\ Phys.\  B {\bf 165}, 483 (1980).

\bibitem{SSW}
  M.~F.~Sohnius, K.~S.~Stelle and P.~C.~West,
 ``Representations of extended supersymmetry,''
in {\it Superspace and Supergravity}, S. W. Hawking and M. Ro\v{c}ek (Eds.), 
Cambridge University Press, Cambridge, 1981, p. 283.  

\bibitem{GS82}
S.~J.~Gates Jr. and W.~Siegel,
``Linearized N=2 superfield supergravity,''  Nucl.\ Phys.\  B {\bf 195}, 39 (1982).

\bibitem{Siegel83}
  W.~Siegel,
  ``Off-shell N=2 supersymmetry for the massive scalar multiplet,''
  Phys.\ Lett.\  B {\bf 122}, 361 (1983).
 
\bibitem{Muller86}
  M.~M\"uller,
  ``Chiral actions for minimal N=2 supergravity,''
  Nucl.\ Phys.\  B {\bf 289}, 557 (1987).


\bibitem{Kuz2008} 
  S.~M.~Kuzenko,
  ``On N = 2 supergravity and projective superspace: Dual formulations,''
  Nucl.\ Phys.\ B {\bf 810}, 135 (2009)
  [arXiv:0807.3381 [hep-th]].

\end{thebibliography}
\end{document}